\documentclass[twoside,twocolumn,9pt]{article}
\usepackage{extsizes}
\usepackage{bm}
\usepackage[super,sort&compress,comma]{natbib} 
\usepackage[version=3]{mhchem}
\usepackage[left=1.5cm, right=1.5cm, top=1.785cm, bottom=2.0cm]{geometry}
\usepackage{balance}
\usepackage{mathptmx}
\usepackage{tabularx}
\usepackage{sectsty}
\usepackage{graphicx} 
\usepackage{lastpage}
\usepackage[format=plain,justification=justified,singlelinecheck=false,font={stretch=1.125,small,sf},labelfont=bf,labelsep=space]{caption}
\usepackage{float}
\usepackage{fancyhdr}
\usepackage{fnpos}
\usepackage[english]{babel}
\addto{\captionsenglish}{%
  
}
\usepackage{array}
\usepackage{droidsans}
\usepackage{charter}
\usepackage[T1]{fontenc}
\usepackage[usenames,dvipsnames]{xcolor}
\usepackage{setspace}
\usepackage[compact]{titlesec}
\usepackage{hyperref}

\usepackage{epstopdf}

\definecolor{cream}{RGB}{222,217,201}

\begin{document}

\pagestyle{fancy}
\thispagestyle{plain}
\fancypagestyle{plain}{
\renewcommand{\headrulewidth}{0pt}
}

\makeFNbottom
\makeatletter
\renewcommand\LARGE{\@setfontsize\LARGE{15pt}{17}}
\renewcommand\Large{\@setfontsize\Large{12pt}{14}}
\renewcommand\large{\@setfontsize\large{10pt}{12}}
\renewcommand\footnotesize{\@setfontsize\footnotesize{7pt}{10}}
\makeatother

\renewcommand{\thefootnote}{\fnsymbol{footnote}}
\renewcommand\footnoterule{\vspace*{1pt}%
\color{cream}\hrule width 3.5in height 0.4pt \color{black}\vspace*{5pt}} 
\setcounter{secnumdepth}{5}

\makeatletter 
\renewcommand\@biblabel[1]{#1}            
\renewcommand\@makefntext[1]%
{\noindent\makebox[0pt][r]{\@thefnmark\,}#1}
\makeatother 
\renewcommand{\figurename}{\small{Fig.}~}
\sectionfont{\sffamily\Large}
\subsectionfont{\normalsize}
\subsubsectionfont{\bf}
\setstretch{1.125} 
\setlength{\skip\footins}{0.8cm}
\setlength{\footnotesep}{0.25cm}
\setlength{\jot}{10pt}
\titlespacing*{\section}{0pt}{4pt}{4pt}
\titlespacing*{\subsection}{0pt}{15pt}{1pt}

\fancyfoot{}
\fancyfoot[LO,RE]{\vspace{-7.1pt}\includegraphics[height=9pt]{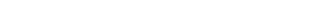}}
\fancyfoot[CO]{\vspace{-7.1pt}\hspace{11.9cm}\includegraphics{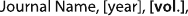}}
\fancyfoot[CE]{\vspace{-7.2pt}\hspace{-13.2cm}\includegraphics{head_foot/RF}}
\fancyfoot[RO]{\footnotesize{\sffamily{1--\pageref{LastPage} ~\textbar  \hspace{2pt}\thepage}}}
\fancyfoot[LE]{\footnotesize{\sffamily{\thepage~\textbar\hspace{4.65cm} 1--\pageref{LastPage}}}}
\fancyhead{}
\renewcommand{\headrulewidth}{0pt} 
\renewcommand{\footrulewidth}{0pt}
\setlength{\arrayrulewidth}{1pt}
\setlength{\columnsep}{6.5mm}
\setlength\bibsep{1pt}

\makeatletter 
\newlength{\figrulesep} 
\setlength{\figrulesep}{0.5\textfloatsep} 

\newcommand{\topfigrule}{\vspace*{-1pt}%
\noindent{\color{cream}\rule[-\figrulesep]{\columnwidth}{1.5pt}} }

\newcommand{\botfigrule}{\vspace*{-2pt}%
\noindent{\color{cream}\rule[\figrulesep]{\columnwidth}{1.5pt}} }

\newcommand{\dblfigrule}{\vspace*{-1pt}%
\noindent{\color{cream}\rule[-\figrulesep]{\textwidth}{1.5pt}} }

\makeatother

\twocolumn[
  \begin{@twocolumnfalse}
{\includegraphics[height=30pt]{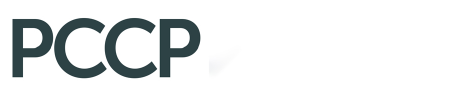}\hfill\raisebox{0pt}[0pt][0pt]{\includegraphics[height=55pt]{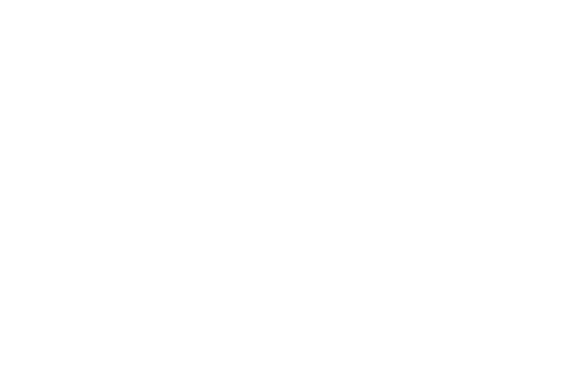}}\\[1ex]
\includegraphics[width=18.5cm]{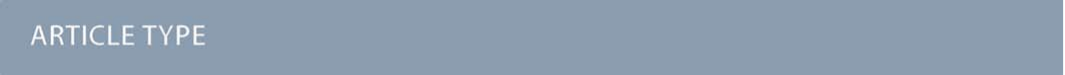}}\par
\vspace{1em}
\sffamily
\begin{tabular}{m{4.5cm} p{13.5cm} }

\includegraphics{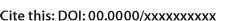} & \noindent\LARGE{\textbf{Optically
    induced spin Hall current in monolayer Janus NbSSe: A
    first-principles study}} \\

\vspace{0.3cm} & \vspace{0.3cm} \\

 & \noindent\large{Souren Adhikary$^1$, Tomoaki Kameda$^1$ and
  Katsunori Wakabayashi$^{1,2,3}$*} \\

\includegraphics{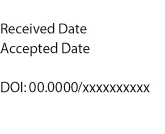} & \noindent\normalsize{Monolayer
  Janus transition-metal dichalcogenides possess Ising- and
  Rashba-type spin-orbit-couplings (SOC), leading to intriguing spin
  splitting effects at K and K$'$, and around $\Gamma$ points across
  the wide energy range. Using first-principles calculations, we
  unveil these SOC characteristics in metallic Janus NbSSe and
  demonstrate its potential for optically controlled spin current
  generation. On the basis of the symmetry of the system, we show that
  different linear polarized light can selectively drive spin currents
  of distinct spin components. Our findings establish NbSSe as a
  promising candidate for next-generation optospintronic technologies,
  which is offering a pathway toward the development of polarization-tunable spin-current sources.}
\\

\end{tabular}

 \end{@twocolumnfalse} \vspace{0.6cm}

]

\renewcommand*\rmdefault{bch}\normalfont\upshape
\rmfamily
\section*{}
\vspace{-1cm}


\footnotetext{\textit{$^{1}$Department of Nanotechnology for Sustainable Energy, School of Science and Technology, Kwansei Gakuin University, 1 Gakuen-Uegahara, Sanda 669-1330, Japan$^1$}}
\footnotetext{\textit{$^{2}$
Center for Spintronics Research Network (CSRN), Osaka University, Toyonaka 560-8531, Japan$^2$ }}

\footnotetext{\textit{$^{3}$ National Institute for Materials Science (NIMS), Namiki 1-1, Tsukuba 305-0044, Japan$^3$}}
\footnotetext{Emails: \textit{sourenadhikary@kwansei.ac.jp, t.kameda@kwansei.ac.jp and  waka@kwansei.ac.jp}}




\section{Introduction}
Conventional electronics are based on charge currents, which generate heat and consume substantial power\cite{appli-1,dyakonov1971current}. In contrast, spin currents can operate on ultra-fast timescales without the need for electrical charge flow, minimizing energy dissipation and allowing faster information processing\cite{rev-1,rev-2,rev-3,rev-4}. The use of light to generate and control spin currents offers a contact-free approach to spin manipulation, reducing device wear, and eliminating interference from electrical contacts\cite{chen2019light,naka2019spin,tao2020pure,gill2025pure}. This advancement paves the way for the development of all-optical spintronic devices and next-generation data storage technologies\cite{appli-2,dyakonov1971possibility,adhikary2021engineering,adhikary2023valley,zhang2024chemically,kaneda2024nanoscrolls}.

Spin-orbit coupling (SOC) plays a crucial role in spin splitting, as well as in the manipulation and detection of spin states. It enables spin-momentum locking and allows for efficient spin polarization of carriers without the need for magnetic ordering. Nonmagnetic materials exhibiting strong SOC are therefore promising candidates for spin-current generation through the spin Hall effect mechanism\cite{dyakonov1971current,dyakonov1971possibility,zhang2000spin,kane2005quantum,avsar2020colloquium,guo2008intrinsic}. Two-dimensional (2D) transition metal dichalcogenides (TMDCs), such as MoS$_2$, WS$_2$, MoSe$_2$, WSe$_2$, exhibit sizable SOC in their occupied valence bands due to the presence of heavy transition metal atoms\cite{xiao2012coupled,feng2012intrinsic,luo2017opto,xu2014spin}. 
The combination of their crystal geometry and broken inversion symmetry leads to SOC-induced lifting of spin degeneracy at the time-reversal symmetric K and K$'$ points: is commonly referred to as Zeeman-type SOC\cite{TMD-3,sanchez2016valley}. This spin splitting at the K and K$'$ valleys occurs in the opposite direction to preserve the time-reversal symmetry. 
Consequently, charge carriers at these points can be distinguished by their spin momentum. This valley-contrasting spin polarization gives rise to a finite spin Berry curvature, and the associated carriers possess finite momentum\cite{feng2012intrinsic, chi2024control,yang2024gate, yang2024twist, li2022designing,chen2022intrinsic}. Therefore, their motion (i.e. spin current) can be controlled using external perturbations such as light. 

\begin{figure*}[h]
\begin{center}
\includegraphics[scale=0.45]{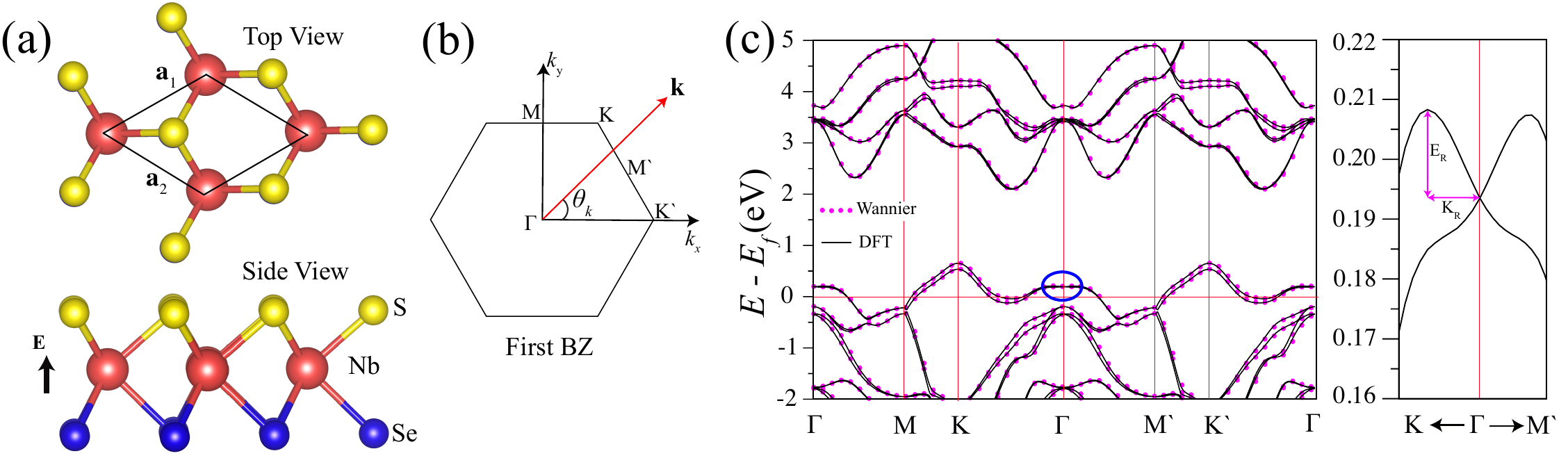}
\caption{(a) The top and side view of lattice structure of monolayer NbSSe within 2H phase. The unit cell of this system show by the rhombus. The shaded areas represent the two layers with two different chalcogen atoms. (b) First BZ and high-symmetric points of NbSSe. (c) The electronic band structure of NbSSe with inclusion of SOC. The vertical (horizontal) dashed lines are the high-symmetric points (Fermi energy). The dotted bands in the band structure is calculating by the Wannier functions. We highlight band splitting (as marked by an ellipse) around $\Gamma$ point in the right panel.}
\label{fig-1}
\end{center}
\end{figure*}

In addition to Zeeman-type SOC, Rashba-type spin–orbit coupling provides an additional mechanism for spin manipulation in 2D materials, as it introduces a linear coupling between the spin and momentum of electrons\cite{yu2021spin,manchon2015new,kapri2021role,chen2022large,patel2022electric,vojacek2024field}. Janus TMDCs are prominent materials for exhibiting Rashba-type SOC, due to the breaking of out-of-plane mirror symmetry, which arises from the presence of two different chalcogen atoms (see Fig. \ref{fig-1}(a))\cite{yao2017manipulation,chakraborty2023anisotropic}. This symmetry breaking induces a built-in out-of-plane electric field \textbf{E}, leading to further spin splitting near the $\Gamma$ point, enhancing spin polarization in these materials\cite{yu2021spin}. As a result, the ability to manipulate in-plane spin polarization using light opens promising opportunities for optospintronic applications. 
Previous studies have primarily focused on semiconducting Rashba-type TMDCs to demonstrate the spin Hall effect\cite{yu2021spin,yang2023tunable,thiruvengadam2022anisotropy,mishra2024preserving}, spin manipulation via optical means in metallic Rashba-type TMDCs remains largely unexplored. Metallic systems offer an advantage over semiconductors in terms of spin Hall conductivity, as they possess a higher density of available charge carriers, which can be tuned more effectively by varying the intensity of the incident light\cite{guo2008intrinsic,wang2024origin}.

Among metallic TMDCs with the 2H phase, monolayer NbSe$_2$ has been extensively studied due to its layer-dependent charge density waves\cite{knispel2024unconventional,xi2015strongly}, quantum metallic state\cite{nakata2021robust,hamill2021two}, and Ising-type superconductivity\cite{wickramaratne2020ising,sohn2018unusual,xi2016ising}. Monolayer NbSe$_2$ has been synthesized either by mechanical exfoliation of bulk NbSe$_2$ or by molecular beam epitaxy\cite{xu2018experimental,de2018tuning}. In monolayer NbSe$_2$, electron spins are pinned in the out-of-plane direction as a result of broken SU(2) spin rotational symmetry, caused by the strong Ising-type spin–orbit interaction. Theoretically, it has been demonstrated that spin currents can be generated in monolayer NbSe$_2$ using linearly polarized light\cite{habara2021optically,habara2022nonlinear,adhikary2024optically}. To introduce Rashba-type SOC, a Janus structure can be engineered from a parent TMDC monolayer by breaking the out-of-plane symmetry. In this study, we focus on the construction of Janus monolayer NbSSe derived from its parent NbSe$_2$. Recently, monolayer NbSSe has been experimentally realized from NbSe$_2$ or NbS$_2$ monolayers using a plasma-assisted chalcogen exchange technique\cite{wu2025metallic,guller2016spin}.

In this study, we investigate the Rashba effect in the Janus NbSSe monolayer using first-principles-based density functional theory (DFT) calculations combined with \textbf{k}.\textbf{p}-based symmetry analysis,
with the aim of exploring its potential for spintronic applications.
We theoretically compute the spin Hall currents corresponding to different spin components under optical irradiation. Our results reveal that a spin current associated with the $x$-component of spin (S$_\textrm{x}$) is generated when $x$- or $y$-polarized light is applied and current measured along the same direction of polarization. In contrast, spin currents associated with the $y$-component of spin (S$_\textrm{y}$) appear only when the current is measured in the transverse direction relative to light polarization (i.e., along the $y$- or $x$-axis for $x$- or $y$-polarized light, respectively). Furthermore, spin currents arising from the $z$-component of spin (S$_\textrm{z}$) consistently exhibit finite values under cross-polarization configurations, where the directions of light polarization and current measurement are orthogonal. Notably, we find that the spin current from the S$_\textrm{y}$ and S$_\textrm{z}$ does not couple with the generated charge current. This indicates the generation of pure spin currents under optical excitation. These findings highlight monolayer NbSSe as a promising material platform for the realization of flexible, two-dimensional spintronic devices.

\begin{figure*}[h]
\begin{center}
\includegraphics[scale=0.6]{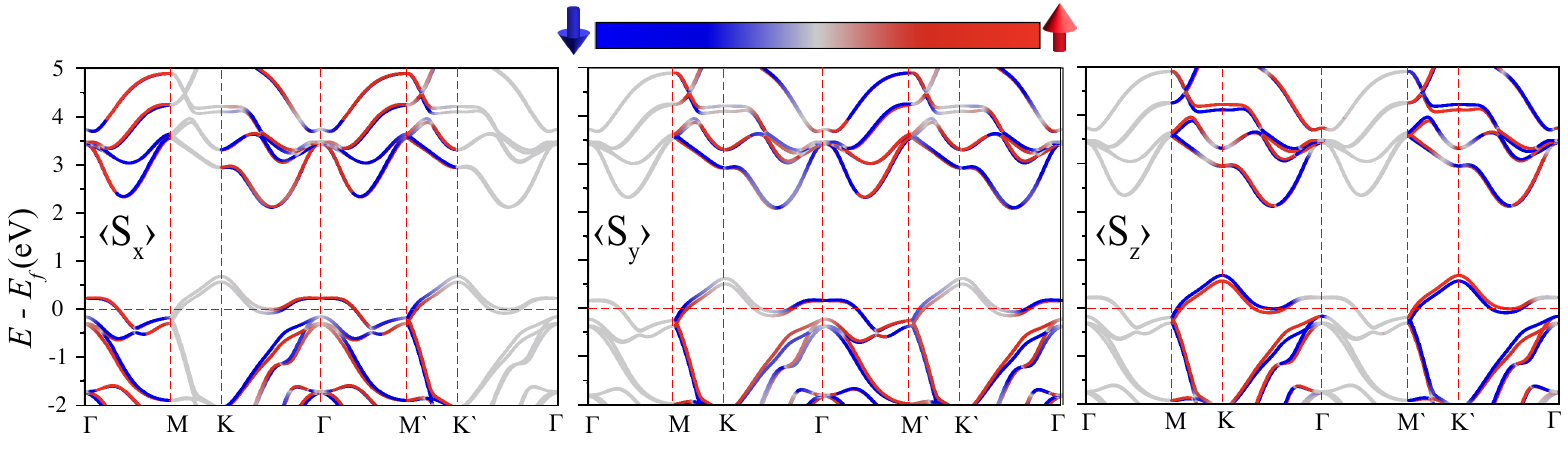}
\caption{The expectation values of three spin components on the electronic band structure of NbSSe. The color bar with red (blue) shows spin up (spin down) orientation.}
\label{fig-2}
\end{center}
\end{figure*}

\section{Methodology and Computational Details}
We calculate the electronic and optical properties of NbSSe monolayer using DFT which is implemented in Quantum Espresso\cite{giannozzi2017advanced}. The electronic exchange-correlation functional is considered within the generalized gradient approximation of Perdew-Burke-Ernzerhof method\cite{perdew1996generalized}. We have used norm-conserving pseudopotentials which are obtained from the Pseudo-Dojo website\cite{van2018pseudodojo}. We have placed the NbSSe monolayer in the $xy$-plane, with a 15 \AA~ vacuum added in the $z$-direction to prevent periodic interactions between adjacent layers. We relaxed the monolayer using the conjugate-gradient method, with a force tolarance of 10$^{-4}$ in atomic units. We set a tolerance of 10$^{-9}$ (in atomic unit) for the electronic relaxation process and the kinetic energy cutoff of the plane wave is 60 Ry. For defining the first Brillouin zone (BZ), we have employed a 16 $\times$ 16 $\times$ 1 $k$-grid. The electronic band structures, including SOC, computed by using a fully relativistic pseudopotential. A Gaussian smearing of 0.005 eV is applied to determine the Fermi energy of the monolayer. To calculate the spin Hall conductivity (SHC) and spin current conductivity (SCC) we construct maximal localized Wannier functions using the Wannier90 package\cite{mostofi2008wannier90,qiao2018calculation,guo2005ab}. Here, we have adopted the d-orbitals of Nb and p-orbitals of S and Se to construct Wannier functions. The optical properties are computed using a converged dense k-grid of 1000 $\times$ 1000 $\times$ 1. Further, to solve equation (\ref{eq-2}), we adopt adaptive smearing method\cite{yates2007spectral}. The state energy difference function i.e., ($E_{nk} - E_{mk}$) is replaced by the $g_{nk}$ ($E_f$) = $\frac{1}{\sqrt{2\pi}W}$exp$(\frac{-(E_f - E_{nk})^2}{2W^2})$, where $W$ is the Gaussian width $E_f$ is the Fermi energy. This width should be, for a given grid spacing $\delta k$, comparable with the energy spacing $\delta E_{nk}$. Commonly, this level of spacing is difficult to estimate due to the flat bands are not properly describe consistently. In adaptive smearing method, level of spacing replaced by a state-depending broadening width  $W_{nm,k} = a |\frac{\partial E_{mk}}{\partial k} - \frac{\partial E_{nk}}{\partial k}| \delta k$. 

\section{Results and Discussions}
\noindent Fig. \ref{fig-1}(a) shows the top and side views of the monolayer NbSSe lattice in the 2H phase. The unit cell is defined by the rhombus and lattice vectors 
\textbf{a}$_1$ and \textbf{a}$_2$. The lattice constant is 3.19 \AA, which is in good agreement with reported experimental values\cite{cossu2020strain,ichinokura2019vortex,wu2025metallic}. The conventional 2H-phase of TMDCs exhibit D$_{3h}$ point-group symmetry, the introduction of two distinct chalcogen atoms in the Janus NbSSe monolayer leads to a reduction in the symmetry to the C$_{3v}$ point-group. This change arises from the inequivalent bond lengths between Nb–S and Nb–Se. As shown in the side view, the presence of S and Se layers on opposite sides of the Nb layer breaks the mirror symmetry in the out-of-plane direction, in addition to the in-plane inversion symmetry. This structure generates an in build electric field (\textbf{E}) in the out-of-plane direction. Fig. \ref{fig-1}(b) illustrates the corresponding first BZ and its high-symmetry points.

To investigate the effect of broken symmetries (i.e., in-plane inversion and out-of-plane mirror); we calculate the electronic band structure, including SOC, and plot it in Fig. \ref{fig-1}(c). The results indicate that NbSSe exhibits metallic behavior, with the Fermi energy (represented by the horizontal red solid line) intersecting several occupied bands. 
Notably, spin splitting is observed around the time-reversal K and K$'$ points, particularly near the Fermi level. The magnitude of spin splitting at the K (or K$'$) points is 115 meV. This spin splitting generally occurs in TMDCs due to Ising- or Zeeman-type SOC\cite{feng2012intrinsic}, which arises from broken in-plane inversion symmetry. In addition, monolayer NbSSe exhibits spin splitting near the $\Gamma$ point (see right panel for the zoomed version), which is characteristic of the Rashba effect. To capture and analyze this Rashba-type splitting, we adopt a \textbf{k}.\textbf{p} based simplified Hamiltonian\cite{manchon2015new}, 
\begin{equation} 
\textrm{H}_\textrm{R} = \alpha_\textrm{R} (k_y\sigma_\textrm{x} -k_x\sigma_\textrm{y})
\label{eq-1}\end{equation}
where $k_x (k_y)$ and $\sigma_\textrm{x} (\sigma_\textrm{y})$ is in-plane electron's momentum in $x$($y$)-direction and Pauli spin matrix, respectively. $\alpha_\textrm{R}$ is the Rashba parameter. The eigenvalues and eigenfunctions of this Hamiltonian are $\pm\alpha_R|\textbf{k}|$ and $|\psi_{\pm}> ~ = \frac{1}{\sqrt{2}}
\begin{pmatrix}
    1  \\
    \mp ie^{i\theta_k} 
\end{pmatrix}$ respectively. Here, $\theta_k$ = tan$^{-1}(k_y/k_x)$ (shown in the Fig. \ref{fig-1}(b)) and $|\textbf{k}|$ = $\sqrt{k_x^2 +k_y^2}$. The Rashba SOC induces energy splitting of $\textrm{E}_\textrm{R}$ and within momentum $\textrm{K}_\textrm{R}$ in the reciprocal space (as shown in Fig. \ref{fig-1}(c)'s right panel). The Rashba parameter is calculated as $\alpha_\textrm{R} = 2\textrm{E}_\textrm{R}/\textrm{K}_\textrm{R}$\cite{zhou2019geometry,sino2021anisotropic}. The calculated  
value of $\alpha_\textrm{R}$ = 384.6 meV$\cdot$\AA~ for the monolayer NbSSe which is higher than semiconducting TMDCs\cite{yu2021spin,yao2017manipulation}. This Rashba SOC can be further increased by applying an external electric field or by stacking layers with a specific order.

To understand spin splitting around the K (or K$'$) and $\Gamma$-point, we calculate the expectation values of the spin components (i.e. $<$~S$_{\textrm{x,y,z}}$~$>$ = $\hbar$/2~$<$$\sigma_{\textrm{x,y,z}}$$>$) in the electronic band structures, using maximal localized Wannier functions based on DFT\cite{mostofi2008wannier90} and present them in Fig. \ref{fig-2}. From Fig. \ref{fig-2}, one can see that the contribution of spin (S$_x$ or S$_y$) depends on the direction of momentum in the BZ. The average value of S$_x$ shows zero contribution along the M-K and K$'$–$\Gamma$ directions, while the average value of S$_y$ exhibits zero contribution along the $\Gamma$–M direction. We attribute this direction-dependent behavior of the in-plane spin components to the symmetry constraints imposed by the Rashba Hamiltonian. We calculate the expectation values of $\sigma_\textrm{x}$ and $\sigma_\textrm{y}$ i.e., $<\psi_+|\sigma_{\textrm{x}}|\psi_+>$ = - sin~$\theta_k$ and $<\psi_+|\sigma_{\textrm{y}}|\psi_+>$ = cos~$\theta_k$ using equation (\ref{eq-1}).  As the average value of S$_y$ depends on 
cos$\theta$, in the $\Gamma$-M direction $\theta$ = 90 i.e.,  cos 90 = 0.  Accordingly, S$_y$ also shows a zero value. Thus, the expectation values of $\sigma_{\textrm{x,y}}$ from our model Hamiltonian and the spin projections obtained from DFT-based calculations show good agreement. However, $<\sigma_\textrm{z}>$, can be obtained without the Rashba SOC because it depends only on the Ising-type SOC\cite{habara2021optically}.
The S$_\textrm{z}$ exhibits significant contributions only near the high-symmetry points K and K$'$. 

The S$_\textrm{x}$ and S$_\textrm{y}$ are contributing around the $\Gamma$ point, whereas the S$_\textrm{z}$ contributes around the K and K$'$ points. Therefore, we have shown that spin splitting occurs in different regions of the BZ, indicating couplings between the electron's spin and momentum. Here, we explore the effect of SOC in the presence of an external perturbation, such as electric field (leading to a spin Hall effect) or the generation of spin currents under optical excitation, which can find optospintronics applications. First, we study this splitting in terms of the spin Hall effect (SHE). We calculate the spin conductivity using the Kubo formula \cite{guo2005ab,qiao2018calculation} as follows:
\begin{figure}[h]
\begin{center}
\includegraphics[scale=0.4]{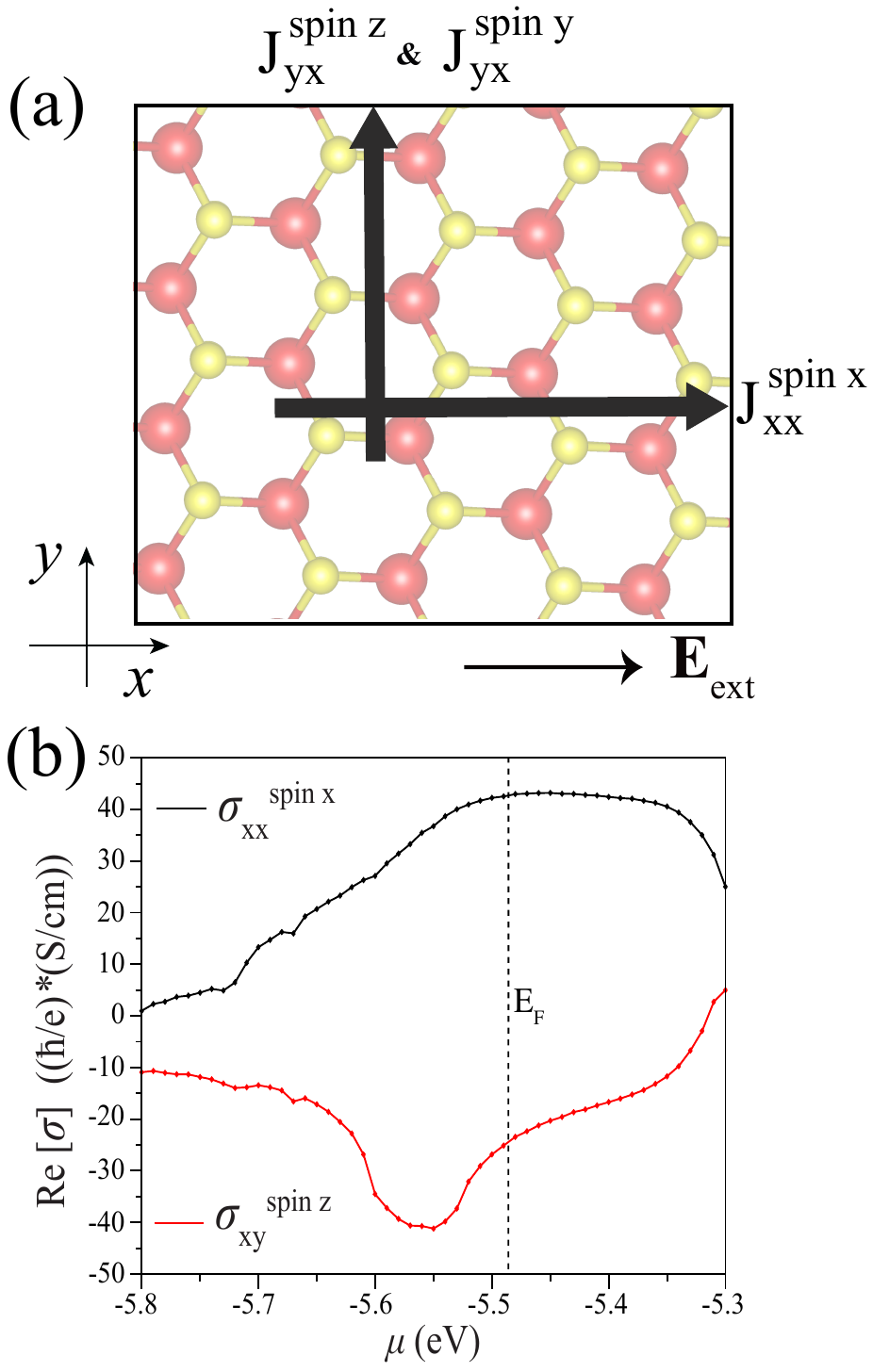}
\caption{(a) The schematic representation of different components of spin current density in presence of an external electric field within. The subscripts denote similar meaning as we defined for spin conductivity in equation \ref{eq-2}. (b) The real part of spin Hall conductivity of monolayer NbSSe with variation of Fermi level in terms of its chemical potential $\mu$. The chemical potential $`$E$_F$' is refer as true Fermi energy of NbSSe.}
\label{fig-3}
\end{center}
\end{figure}
\begin{equation}
\sigma_{\alpha\beta}^{\textrm{spin}~ \gamma} = \frac{\textrm{e}^2}{\hbar}\frac{\hbar^2}{\textrm{V}N_k}\sum_k \sum_n f_{n\textbf{k}}
 \sum_{m\neq n} \frac{-2\textrm{Im} [<n\textbf{k}|\hat{j}^{\gamma}_{\alpha}|m\textbf{k}><m\textbf{k}|\hat{\nu}_{\beta}|n\textbf{k}>]}{(E_{nk} - E_{mk})^2 - (\hbar\omega + i\eta)^2}
\label{eq-2}\end{equation}
where $n(m)$ is the band indices including spin degree of freedom, V is the primitive cell volume, $N_k$ is the number of k-points used for sampling the BZ, $E_{nk}$($E_{mk}$) are the eigenvalues, $f_{nk}$ is the Fermi distribution function and $\omega$ is the frequency of incident light. The terms $\hbar$ and $e$ are the reduced Plank constant and electronic charge of the electron, respectively. The spin current operator is written as  $\hat{j}^{\gamma}_{\alpha} = \frac{1}{2}\{\hat{s}_{\gamma},\hat{\nu}_{\alpha}\}$, $\hat{s}_{\gamma}$ is spin operator and $\hat{\nu}_{\alpha,\beta}$ is the velocity operator of electron in ($\alpha,\beta$) direction. The quantity $\alpha$ or $\beta$
can take $x$ or $y$-direction and $\gamma$ can be $x$,$y$ or $z$-component of electron's spin. In the context of the spin Hall effect, 
$\alpha$ represents the direction of spin movement, while $\beta$ denotes the direction of the applied electric field or gate voltage. That is, if a gate voltage or electric field is applied in the $\beta$ direction, a transverse spin current will be generated in the $\alpha$ direction. 

Here, we first analyze the dependence of SHC on the direction of the applied electric field ($\beta$) and the movement of spins ($\alpha$).  Since the electron spin operator is a pseudovector and the velocity operator is a linear operator, their combination results in a third-rank pseudotensor in spin-dependent conductivity $\sigma^{\textrm{spin}~\gamma}_{\alpha\beta}$. Therefore, $\sigma^{\textrm{spin}~\gamma}_{\alpha\beta}$ will have four components for each value of $\gamma$ (as NbSSe is a 2D material then $\alpha$ ($\beta$) can take two values i.e., $x$ or $y$). The general expression of spin Hall conductivity for different spin components is as follows:
\begin{align*}
\sigma^{\textrm{spin}~x}=\begin{pmatrix}
    \sigma^{\textrm{spin} ~x}_{xx} &\sigma^{\textrm{spin} ~x}_{xy}\\
    \sigma^{\textrm{spin} ~x}_{yx}&\sigma^{\textrm{spin} ~x}_{yy}\\
    \end{pmatrix},
    \end{align*}
    \begin{align*}
  \sigma^{\textrm{spin}~y}=  \begin{pmatrix}
    \sigma^{\textrm{spin} ~y}_{xx} &\sigma^{\textrm{spin} ~y}_{xy} \\
    \sigma^{\textrm{spin} ~y}_{yx}&\sigma^{\textrm{spin} ~y}_{yy}
    \end{pmatrix},
    \end{align*}
    \begin{align}
  \sigma^{\textrm{spin}~z} =   \begin{pmatrix}
    \sigma^{\textrm{spin} ~z}_{xx} &\sigma^{\textrm{spin} ~z}_{xy} \\
    \sigma^{\textrm{spin} ~z}_{yx}&\sigma^{\textrm{spin} ~z}_{yy}
    \end{pmatrix}.
    \end{align}
From our DFT based calculations, we find $\sigma_{xx}^{\textrm{spin}~ x}$ = - $\sigma_{yy}^{\textrm{spin}~ x}$ and $\sigma_{xy}^{\textrm{spin}~ x}$ = $\sigma_{yx}^{\textrm{spin}~ x}$ = 0 for S$_\textrm{x}$. The S$_\textrm{y}$ shows $\sigma_{xx}^{\textrm{spin}~ y}$ = $\sigma_{yy}^{\textrm{spin}~ y}$ = 0 and 
$\sigma_{xy}^{\textrm{spin}~ y}$ = $\sigma_{yx}^{\textrm{spin}~ y}$ = - $\sigma_{xx}^{\textrm{spin}~ x}$. In the case of S$_\textrm{z}$, $\sigma_{xx}^{\textrm{spin}~ z}$ = $\sigma_{yy}^{\textrm{spin}~ z}$ = 0, $\sigma_{xy}^{\textrm{spin}~ z}$ and $\sigma_{yx}^{\textrm{spin}~ z}$ has finite value with $\sigma_{xy}^{\textrm{spin}~ z}$ = - $\sigma_{yx}^{\textrm{spin}~ z}$. Our calculated results for the different components of the spin conductivity are consistent with the symmetry-based Neumann’s principle for Janus TMDCs\cite{seemann2015symmetry,xiang2024classification}.
In summary, a finite spin current will be generate due to the S$_\textrm{x}$ is observed when the electric field is applied in $x ~(y)$ and the current is measured in the same $x~(y)$-direction. On the other hand, th S$_\textrm{y}$ and S$_\textrm{z}$ shows finite value only in cross configurations of applied electric field and spin movement. We show a schematic representation of spin Hall conductivity (in terms of spin current density) for different spin components when applied an external electric field (\textbf{E}$_{\textrm{ext}}$) in $x$ direction in Fig. \ref{fig-3}(a). Furthermore, we plot only the $\sigma_{xx}^{\textrm{spin}~ x}$ and 
$\sigma_{xy}^{\textrm{spin}~ z}$ in Fig. \ref{fig-3}(b), while the remaining components can be obtained using the relations mentioned above. We vary the Fermi level (since the Fermi function depends on the chemical potential of the system) in equation (\ref{eq-2}) and plot direct current (DC) SHC (i.e., in the clean limit $\omega$ = 0) in Fig. \ref{fig-3}. Here, $\eta$ considers via the ``adaptive smearing'' scheme \cite{yates2007spectral}.  The value of the spin Hall conductivity at the Fermi level is comparable to that of monolayer NbSe$_2$ \cite{adhikary2024optically} and higher than that of the MoSSe system\cite{yu2021spin}. These generated spin currents can be measured experimentally using Spin-Resolved Photoemission Spectroscopy (SR-PES) or Spin-Polarized Scanning Tunneling Microscopy (SP-STM)\cite{kubetzka2002spin}. In SR-PES, spin polarization components can be detected by measuring intensity variations\cite{kawaguchi2023time}. Similarly, in SP-STM, a magnetized tip is used to probe the local spin density in different spatial directions. By changing the magnetization direction of the tip, different spin components can be analyzed through variations in the tunneling current, which depends on the spin orientation of the sample.

\begin{figure}[h]
\begin{center}
\includegraphics[scale=0.55]{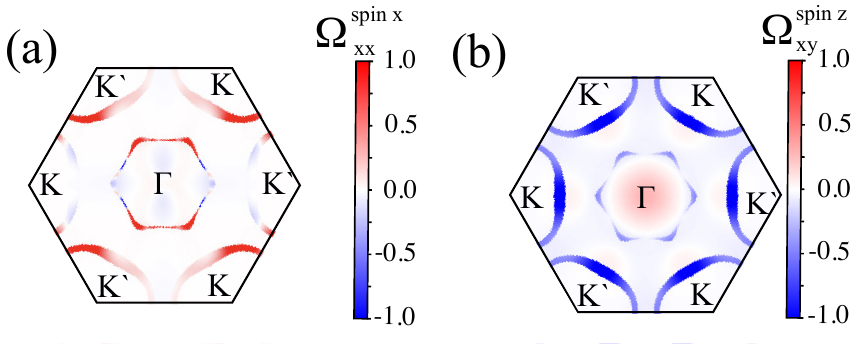}
\caption{The spin Berry curvature (SBC) for (a) S$_\textrm{x}$ (b) S$_\textrm{z}$ of NbSSe. Here we plot SBC in a log scale via $\Omega$ = sgn(SBC)log$_{10}$$|$SBC$|$ if $|$SBC$|$ $>$ 10 or  $\Omega$ = SBC/10 if $|$SBC$|$ $\leq$ 10.  Where, sgn(SBC) means taking the sign of SBC. Finally, we normalize the SBC with respect to its maximum value and plot it within +1 to -1.}
\label{fig-4}
\end{center}
\end{figure}

Next, to determine the origin of this spin current, we calculate the spin Berry curvature (SBC) using Kubo formula as follows \cite{guo2005ab,qiao2018calculation}:
\begin{equation}
\Omega_{\alpha\beta}^{\textrm{spin}~ \gamma} = -\hbar^2\sum_n f_{n\textbf{k}}
 \sum_{m\neq n} \frac{2\textrm{Im} [<n\textbf{k}|\hat{j}^{\gamma}_{\alpha}|m\textbf{k}><m\textbf{k}|\hat{\nu}_{\beta}|n\textbf{k}>]}{(E_{nk} - E_{mk})^2}
\label{eq-4}\end{equation}
where, different quantities in this equation we have already defined in  equation (\ref{eq-2}). Here we set the energy eigenvalue at the true Fermi energy of NbSSe. We plot $\Omega_{xx}^{\textrm{spin}~ x}$ and $\Omega_{xy}^{\textrm{spin}~ z}$ in Fig. \ref{fig-4}(a) \& (b), respectively. The SBC for S$_\textrm{z}$ is showing C$_6$ symmetry due to the preserved time-reversal symmetry. On the other hand, SBC due to the S$_\textrm{x}$ maintains the angle dependence similar to the S$_\textrm{x}$ contribution in the spin splitting (see Fig. \ref{fig-2}). In particular, both the SBCs, $\Omega_{xy}^{\textrm{spin}~ z}$ and $\Omega_{xx}^{\textrm{spin}~ x}$ exhibit a total finite value throughout the BZ. Thus, the finite values of SBC can act as an effective magnetic field in reciprocal space\cite{kane2005quantum}. As a consequence, spins can move with anomalous velocity (\textbf{v}), leading to the generation of spin current by applying the electric field (\textbf{E}$_{\textrm{ext}}$) via $\textbf{v} = -\frac{e}{\hbar} \textbf{E}_{\textrm{ext}} \times \bm{\Omega}$. These results reveal that the effects of Rashba SOC in monolayer NbSSe can generate a spin current with in-plane components (i.e., S$_\textrm{x}$ and S$_\textrm{y}$) in addition to the out-of-plane (S$_\textrm{z}$) spin current. Thus, monolayer NbSSe holds significant potential for applications in spintronics devices. Next, we explore this spin current generation using an external green energy source, i.e., linear polarized light. 

\begin{figure}[h]
\begin{center}
\includegraphics[scale=0.5]{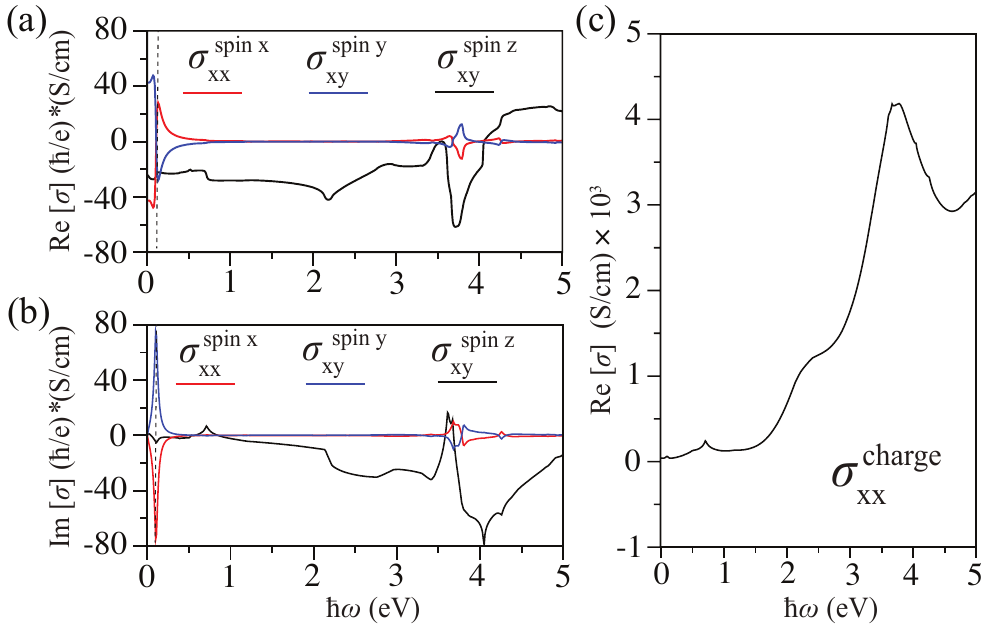}
\caption{The angular frequency dependent spin current (a) real part , (b) imaginary part of different spin components of NbSSe with presence of light. The vertical dashed line represents the Rashba position. (c) Charge current generation on NbSSe due to the $x$-polarized light in $x$-direction.}
\label{fig-5}
\end{center}
\end{figure}

We calculate the frequency-dependent spin-conductivity by the using equation \ref{eq-2} at the true Fermi energy of NbSSe. To set the parameter $\eta$ we have used adaptive smearing method\cite{yates2007spectral,qiao2018calculation}. We plot the real and imaginary part of frequency-dependent spin conductivity for the S$_\textrm{x}$, S$_\textrm{y}$, and S$_\textrm{z}$ in Fig. \ref{fig-5}(a)\&(b), respectively. The spin current due to S$_\textrm{x}$ shows a finite value when $x$-polarized light is applied and current is measured in the 
$x$-direction. Similarly, the spin current due to S$_\textrm{y}$ can be generated in the transverse (i.e., $x$) direction using $y$-polarized light. We observe that both the S$_\textrm{x}$ and S$_\textrm{y}$ current exhibit finite values when the frequency of light is zero, corresponding to the DC limit (i.e., $\omega$ = 0). As the frequency of light increases, a peak appears at 0.1 eV of photon energy for both S$_\textrm{x}$ and S$_\textrm{y}$ current. This peak corresponds to the energy where Rashba spin splitting occurs in the electronic band structure, which is approximately 0.1 eV above the Fermi level (see Fig. \ref{fig-1}(c)). Beyond this point, both spin currents drop to zero. These spin current conductivities are comparable to those observed in GaAs\cite{qiao2018calculation}, 2D InSe\cite{farooq2022spin} and monolayer MoSSe\cite{yu2021spin}.
In addition to the in-plane spin-polarized current, the out-of-plane spin-polarized current can be generated using light, as monolayer NbSSe exhibits Ising-type SOC. Therefore, in-plane as well as out-of-plane spin polarized currents can be generated with a wide range of light energy, highlighting the potential applications of monolayer NbSSe in optospintronics. The imaginary part of these optical conductivities represents the amount of energy storage during this optical process.

Further, we check the purity of these spin currents by calculating the charge current generated by light with different linear polarizations. The optical conductivity due to charge is obtained by replacing the spin-current operator $\hat{j}^{\gamma}_{\alpha} = \frac{1}{2}\{\hat{s}_{\gamma},\hat{\nu}_{\alpha}\}$ by the velocity operator $\hat{\nu}_{\alpha}$ in equation (\ref{eq-2}). We plot the charge current under light irradiation in Fig. \ref{fig-5}(c). A finite charge current is observed in configurations where the direction of polarization of applied light and the measurement direction are the same. Specifically, a finite charge current occurs when $x$ (or $y$)-polarized light is applied, and the current is measured in the same $x$ (or $y$)-direction i.e., $\sigma_{xx}^{\textrm{charge}}$ = $\sigma_{yy}^{\textrm{charge}}$ shows finite values. Here, we have plotted only the $\sigma_{xx}^{\textrm{charge}}$. The cross configurations show zero charge current, which means that when $x$-polarized light is applied, no charge current is generated in the transverse $y$-direction, and vice versa (i.e., $\sigma_{xy}^{\textrm{charge}}$ = $\sigma_{yx}^{\textrm{charge}}$ = 0). From this study, we conclude that the spin currents associated with S$_\textrm{y}$ and S$_\textrm{z}$ are pure in nature, as they do not produce a simultaneous charge current. However, the spin current due to S$_\textrm{x}$ is mixed, which means that both spin and charge currents are generated simultaneously. This mixed generation of spin and charge currents is less desirable due to contamination from Joule heating and rectification effects. Therefore, a pure spin current is more desirable for high-speed spin-photovoltaic devices. Note that the range of these optically generated charge currents is comparable to that of other metallic systems\cite{gong2024large,homes2013optical,schindlmayr2009optical}.

\section*{Conclusions}
In conclusion, using first-principles calculations, we have explored the optically controlled spintronic properties of metallic NbSSe monolayer. Monolayer NbSSe exhibits sizable SOC strength due to the presence of Nb. This system shares a hexagonal structure with broken in-plane symmetry and demonstrates Ising-type opposite spin splitting at the time-reversal K and K$'$ points, resulting in electron spins oriented in the out-of-plane direction. Additionally, owing to  the broken out-of-plane mirror symmetry, this system exhibits Rashba SOC, enabling the in-plane spin orientations.  These spin splittings and preserve time-reversal symmetry, monolayer NbSSe exhibits a finite SBC in whole BZ. We have investigated this finite SBC in the presence of linear polarized light to induced spin currents.  We found that linearly polarized light (with a certain polarization) can induce spin currents associated with the three components of the electron's spin (S$_\textrm{x}$, S$_\textrm{y}$ and S$_\textrm{z}$). However, the spin currents associated with S$_\textrm{y}$ and S$_\textrm{z}$ are induced in the direction transverse to the light's polarization, whereas the current associated with S$_\textrm{x}$ is induced in the same direction as the light's polarization. Further, we have revealed that the spin currents associated with S$_\textrm{y}$ and S$_\textrm{z}$ are pure, whereas the spin current associated with S$_\textrm{x}$  is mixed with a charge current due to the symmetry. Therefore, monolayer NbSSe holds great potential for applications in optospintronic devices. In general, light-induced spin currents can be measured through the inverse spin Hall effect (ISHE). When one edge of the sample accumulates spin (due to conservation of total angular momentum), a conduction-electron spin current ($J_s$) can be generated\cite{wang2024origin}. This spin current can then be detected as a voltage via ISHE. However, the intensity of the incident light produces different effects on spin-current generation. For example, high-intensity light can generate thermally excited spin currents, which in turn affect the mean free path of the light-induced spin current. Additionally, spin-flop scattering (spin life time) caused by impurities, vacancies, finite temperature, and defects reduces the outgoing spin angular momentum.

\section*{Author contributions}
Souren Adhikary: Formal analysis, Investigation, Data curation,
Writing – original draft, Visualization, Software. Tomoaki Kameda: Investigation, Formal analysis. Katsunori Wakabayashi: Conceptualization, Methodology, Project
administration, Supervision, Writing –review \& editing.

\section*{Conflicts of interest}
There are no conflicts to declare.

\section*{Data availability}
Additional data or computational files are available from the corresponding author upon reasonable request.
\section*{Acknowledgements}
This work was supported by JSPS KAKENHI (Grants No. JP25K01609,
No. JP22H05473, and No. JP21H01019), JST CREST (Grant
No. JPMJCR19T1). K. W. acknowledges the financial support for Basic
Science Research Projects (Grant No. 2401203) from the Sumitomo Foundation.



\balance


\bibliography{rsc} 

\providecommand*{\mcitethebibliography}{\thebibliography}
\csname @ifundefined\endcsname{endmcitethebibliography}
{\let\endmcitethebibliography\endthebibliography}{}
\begin{mcitethebibliography}{76}
\providecommand*{\natexlab}[1]{#1}
\providecommand*{\mciteSetBstSublistMode}[1]{}
\providecommand*{\mciteSetBstMaxWidthForm}[2]{}
\providecommand*{\mciteBstWouldAddEndPuncttrue}
  {\def\EndOfBibitem{\unskip.}}
\providecommand*{\mciteBstWouldAddEndPunctfalse}
  {\let\EndOfBibitem\relax}
\providecommand*{\mciteSetBstMidEndSepPunct}[3]{}
\providecommand*{\mciteSetBstSublistLabelBeginEnd}[3]{}
\providecommand*{\EndOfBibitem}{}
\mciteSetBstSublistMode{f}
\mciteSetBstMaxWidthForm{subitem}
{(\emph{\alph{mcitesubitemcount}})}
\mciteSetBstSublistLabelBeginEnd{\mcitemaxwidthsubitemform\space}
{\relax}{\relax}

\bibitem[{\v{Z}}uti{\'c} \emph{et~al.}(2004){\v{Z}}uti{\'c}, Fabian, and
  Sarma]{appli-1}
I.~{\v{Z}}uti{\'c}, J.~Fabian and S.~D. Sarma, \emph{Rev. Mod. Phys.}, 2004,
  \textbf{76}, 323\relax
\mciteBstWouldAddEndPuncttrue
\mciteSetBstMidEndSepPunct{\mcitedefaultmidpunct}
{\mcitedefaultendpunct}{\mcitedefaultseppunct}\relax
\EndOfBibitem
\bibitem[Dyakonov and Perel(1971)]{dyakonov1971current}
M.~I. Dyakonov and V.~Perel, \emph{Phys. Lett. A}, 1971, \textbf{35},
  459--460\relax
\mciteBstWouldAddEndPuncttrue
\mciteSetBstMidEndSepPunct{\mcitedefaultmidpunct}
{\mcitedefaultendpunct}{\mcitedefaultseppunct}\relax
\EndOfBibitem
\bibitem[Jungwirth \emph{et~al.}(2012)Jungwirth, Wunderlich, and
  Olejn{\'\i}k]{rev-1}
T.~Jungwirth, J.~Wunderlich and K.~Olejn{\'\i}k, \emph{Nat. Mater.}, 2012,
  \textbf{11}, 382--390\relax
\mciteBstWouldAddEndPuncttrue
\mciteSetBstMidEndSepPunct{\mcitedefaultmidpunct}
{\mcitedefaultendpunct}{\mcitedefaultseppunct}\relax
\EndOfBibitem
\bibitem[Hirsch(1999)]{rev-2}
J.~Hirsch, \emph{Phys. Rev. Lett.}, 1999, \textbf{83}, 1834\relax
\mciteBstWouldAddEndPuncttrue
\mciteSetBstMidEndSepPunct{\mcitedefaultmidpunct}
{\mcitedefaultendpunct}{\mcitedefaultseppunct}\relax
\EndOfBibitem
\bibitem[Sinova \emph{et~al.}(2004)Sinova, Culcer, Niu, Sinitsyn, Jungwirth,
  and MacDonald]{rev-3}
J.~Sinova, D.~Culcer, Q.~Niu, N.~Sinitsyn, T.~Jungwirth and A.~H. MacDonald,
  \emph{Phys. Rev. Lett.}, 2004, \textbf{92}, 126603\relax
\mciteBstWouldAddEndPuncttrue
\mciteSetBstMidEndSepPunct{\mcitedefaultmidpunct}
{\mcitedefaultendpunct}{\mcitedefaultseppunct}\relax
\EndOfBibitem
\bibitem[Valenzuela and Tinkham(2006)]{rev-4}
S.~O. Valenzuela and M.~Tinkham, \emph{Nature}, 2006, \textbf{442},
  176--179\relax
\mciteBstWouldAddEndPuncttrue
\mciteSetBstMidEndSepPunct{\mcitedefaultmidpunct}
{\mcitedefaultendpunct}{\mcitedefaultseppunct}\relax
\EndOfBibitem
\bibitem[Chen and Huang(2019)]{chen2019light}
Y.-J. Chen and S.-Y. Huang, \emph{Phys. Rev. B}, 2019, \textbf{99},
  094426\relax
\mciteBstWouldAddEndPuncttrue
\mciteSetBstMidEndSepPunct{\mcitedefaultmidpunct}
{\mcitedefaultendpunct}{\mcitedefaultseppunct}\relax
\EndOfBibitem
\bibitem[Naka \emph{et~al.}(2019)Naka, Hayami, Kusunose, Yanagi, Motome, and
  Seo]{naka2019spin}
M.~Naka, S.~Hayami, H.~Kusunose, Y.~Yanagi, Y.~Motome and H.~Seo, \emph{Nat.
  Commun.}, 2019, \textbf{10}, 4305\relax
\mciteBstWouldAddEndPuncttrue
\mciteSetBstMidEndSepPunct{\mcitedefaultmidpunct}
{\mcitedefaultendpunct}{\mcitedefaultseppunct}\relax
\EndOfBibitem
\bibitem[Tao \emph{et~al.}(2020)Tao, Jiang, Hao, Zheng, Zhang, and
  Zeng]{tao2020pure}
X.~Tao, P.~Jiang, H.~Hao, X.~Zheng, L.~Zhang and Z.~Zeng, \emph{Phys. Rev. B},
  2020, \textbf{102}, 081402\relax
\mciteBstWouldAddEndPuncttrue
\mciteSetBstMidEndSepPunct{\mcitedefaultmidpunct}
{\mcitedefaultendpunct}{\mcitedefaultseppunct}\relax
\EndOfBibitem
\bibitem[Gill \emph{et~al.}(2025)Gill, Sharma, and Shallcross]{gill2025pure}
D.~Gill, S.~Sharma and S.~Shallcross, \emph{Nano Lett.}, 2025\relax
\mciteBstWouldAddEndPuncttrue
\mciteSetBstMidEndSepPunct{\mcitedefaultmidpunct}
{\mcitedefaultendpunct}{\mcitedefaultseppunct}\relax
\EndOfBibitem
\bibitem[Wolf \emph{et~al.}(2001)Wolf, Awschalom, Buhrman, Daughton, von
  Moln{\'a}r, Roukes, Chtchelkanova, and Treger]{appli-2}
S.~Wolf, D.~Awschalom, R.~Buhrman, J.~Daughton, v.~S. von Moln{\'a}r,
  M.~Roukes, A.~Y. Chtchelkanova and D.~Treger, \emph{Science}, 2001,
  \textbf{294}, 1488--1495\relax
\mciteBstWouldAddEndPuncttrue
\mciteSetBstMidEndSepPunct{\mcitedefaultmidpunct}
{\mcitedefaultendpunct}{\mcitedefaultseppunct}\relax
\EndOfBibitem
\bibitem[Dyakonov(1971)]{dyakonov1971possibility}
M.~I. Dyakonov, \emph{JETP Lett.}, 1971, \textbf{13}, 467\relax
\mciteBstWouldAddEndPuncttrue
\mciteSetBstMidEndSepPunct{\mcitedefaultmidpunct}
{\mcitedefaultendpunct}{\mcitedefaultseppunct}\relax
\EndOfBibitem
\bibitem[Adhikary \emph{et~al.}(2021)Adhikary, Mohakud, and
  Dutta]{adhikary2021engineering}
S.~Adhikary, S.~Mohakud and S.~Dutta, \emph{Phys. Status Solidi b}, 2021,
  \textbf{258}, 2100071\relax
\mciteBstWouldAddEndPuncttrue
\mciteSetBstMidEndSepPunct{\mcitedefaultmidpunct}
{\mcitedefaultendpunct}{\mcitedefaultseppunct}\relax
\EndOfBibitem
\bibitem[Adhikary \emph{et~al.}(2023)Adhikary, Mohakud, and
  Dutta]{adhikary2023valley}
S.~Adhikary, S.~Mohakud and S.~Dutta, \emph{Phys. Rev. B}, 2023, \textbf{108},
  195429\relax
\mciteBstWouldAddEndPuncttrue
\mciteSetBstMidEndSepPunct{\mcitedefaultmidpunct}
{\mcitedefaultendpunct}{\mcitedefaultseppunct}\relax
\EndOfBibitem
\bibitem[Zhang \emph{et~al.}(2024)Zhang, Liu, Nakajo, Aoki, Wang, Wang, Gao,
  Maruyama, Kawakami, Makino,\emph{et~al.}]{zhang2024chemically}
W.~Zhang, Z.~Liu, H.~Nakajo, S.~Aoki, H.~Wang, Y.~Wang, Y.~Gao, M.~Maruyama,
  T.~Kawakami, Y.~Makino \emph{et~al.}, \emph{Small Structures}, 2024,
  \textbf{5}, 2300514\relax
\mciteBstWouldAddEndPuncttrue
\mciteSetBstMidEndSepPunct{\mcitedefaultmidpunct}
{\mcitedefaultendpunct}{\mcitedefaultseppunct}\relax
\EndOfBibitem
\bibitem[Kaneda \emph{et~al.}(2024)Kaneda, Zhang, Liu, Gao, Maruyama,
  Nakanishi, Nakajo, Aoki, Honda, Ogawa,\emph{et~al.}]{kaneda2024nanoscrolls}
M.~Kaneda, W.~Zhang, Z.~Liu, Y.~Gao, M.~Maruyama, Y.~Nakanishi, H.~Nakajo,
  S.~Aoki, K.~Honda, T.~Ogawa \emph{et~al.}, \emph{ACS Nano}, 2024,
  \textbf{18}, 2772--2781\relax
\mciteBstWouldAddEndPuncttrue
\mciteSetBstMidEndSepPunct{\mcitedefaultmidpunct}
{\mcitedefaultendpunct}{\mcitedefaultseppunct}\relax
\EndOfBibitem
\bibitem[Zhang(2000)]{zhang2000spin}
S.~Zhang, \emph{Phys. Rev. Lett.}, 2000, \textbf{85}, 393\relax
\mciteBstWouldAddEndPuncttrue
\mciteSetBstMidEndSepPunct{\mcitedefaultmidpunct}
{\mcitedefaultendpunct}{\mcitedefaultseppunct}\relax
\EndOfBibitem
\bibitem[Kane and Mele(2005)]{kane2005quantum}
C.~L. Kane and E.~J. Mele, \emph{Phys. Rev. Lett.}, 2005, \textbf{95},
  226801\relax
\mciteBstWouldAddEndPuncttrue
\mciteSetBstMidEndSepPunct{\mcitedefaultmidpunct}
{\mcitedefaultendpunct}{\mcitedefaultseppunct}\relax
\EndOfBibitem
\bibitem[Avsar \emph{et~al.}(2020)Avsar, Ochoa, Guinea, {\"O}zyilmaz, Van~Wees,
  and Vera-Marun]{avsar2020colloquium}
A.~Avsar, H.~Ochoa, F.~Guinea, B.~{\"O}zyilmaz, B.~Van~Wees and I.~J.
  Vera-Marun, \emph{Rev. Mod. Phys.}, 2020, \textbf{92}, 021003\relax
\mciteBstWouldAddEndPuncttrue
\mciteSetBstMidEndSepPunct{\mcitedefaultmidpunct}
{\mcitedefaultendpunct}{\mcitedefaultseppunct}\relax
\EndOfBibitem
\bibitem[Guo \emph{et~al.}(2008)Guo, Murakami, Chen, and
  Nagaosa]{guo2008intrinsic}
G.-Y. Guo, S.~Murakami, T.-W. Chen and N.~Nagaosa, \emph{Phys. Rev. Lett.},
  2008, \textbf{100}, 096401\relax
\mciteBstWouldAddEndPuncttrue
\mciteSetBstMidEndSepPunct{\mcitedefaultmidpunct}
{\mcitedefaultendpunct}{\mcitedefaultseppunct}\relax
\EndOfBibitem
\bibitem[Xiao \emph{et~al.}(2012)Xiao, Liu, Feng, Xu, and Yao]{xiao2012coupled}
D.~Xiao, G.-B. Liu, W.~Feng, X.~Xu and W.~Yao, \emph{Phys. Rev. Lett.}, 2012,
  \textbf{108}, 196802\relax
\mciteBstWouldAddEndPuncttrue
\mciteSetBstMidEndSepPunct{\mcitedefaultmidpunct}
{\mcitedefaultendpunct}{\mcitedefaultseppunct}\relax
\EndOfBibitem
\bibitem[Feng \emph{et~al.}(2012)Feng, Yao, Zhu, Zhou, Yao, and
  Xiao]{feng2012intrinsic}
W.~Feng, Y.~Yao, W.~Zhu, J.~Zhou, W.~Yao and D.~Xiao, \emph{Phys. Rev. B},
  2012, \textbf{86}, 165108\relax
\mciteBstWouldAddEndPuncttrue
\mciteSetBstMidEndSepPunct{\mcitedefaultmidpunct}
{\mcitedefaultendpunct}{\mcitedefaultseppunct}\relax
\EndOfBibitem
\bibitem[Luo \emph{et~al.}(2017)Luo, Xu, Zhu, Wu, McCormick, Zhan, Neupane, and
  Kawakami]{luo2017opto}
Y.~K. Luo, J.~Xu, T.~Zhu, G.~Wu, E.~J. McCormick, W.~Zhan, M.~R. Neupane and
  R.~K. Kawakami, \emph{Nano Lett.}, 2017, \textbf{17}, 3877--3883\relax
\mciteBstWouldAddEndPuncttrue
\mciteSetBstMidEndSepPunct{\mcitedefaultmidpunct}
{\mcitedefaultendpunct}{\mcitedefaultseppunct}\relax
\EndOfBibitem
\bibitem[Xu \emph{et~al.}(2014)Xu, Yao, Xiao, and Heinz]{xu2014spin}
X.~Xu, W.~Yao, D.~Xiao and T.~F. Heinz, \emph{Nat. Phys.}, 2014, \textbf{10},
  343--350\relax
\mciteBstWouldAddEndPuncttrue
\mciteSetBstMidEndSepPunct{\mcitedefaultmidpunct}
{\mcitedefaultendpunct}{\mcitedefaultseppunct}\relax
\EndOfBibitem
\bibitem[Cao \emph{et~al.}(2012)Cao, Wang, Han, Ye, Zhu, Shi, Niu, Tan, Wang,
  Liu,\emph{et~al.}]{TMD-3}
T.~Cao, G.~Wang, W.~Han, H.~Ye, C.~Zhu, J.~Shi, Q.~Niu, P.~Tan, E.~Wang, B.~Liu
  \emph{et~al.}, \emph{Nat. Commun.}, 2012, \textbf{3}, 887\relax
\mciteBstWouldAddEndPuncttrue
\mciteSetBstMidEndSepPunct{\mcitedefaultmidpunct}
{\mcitedefaultendpunct}{\mcitedefaultseppunct}\relax
\EndOfBibitem
\bibitem[Sanchez \emph{et~al.}(2016)Sanchez, Ovchinnikov, Misra, Allain, and
  Kis]{sanchez2016valley}
O.~L. Sanchez, D.~Ovchinnikov, S.~Misra, A.~Allain and A.~Kis, \emph{Nano
  Lett.}, 2016, \textbf{16}, 5792--5797\relax
\mciteBstWouldAddEndPuncttrue
\mciteSetBstMidEndSepPunct{\mcitedefaultmidpunct}
{\mcitedefaultendpunct}{\mcitedefaultseppunct}\relax
\EndOfBibitem
\bibitem[Chi \emph{et~al.}(2024)Chi, Lee, Yang, Dolan, Safeer, Ingla-Ayn{\'e}s,
  Herling, Ontoso, Mart{\'\i}n-Garc{\'\i}a,
  Gobbi,\emph{et~al.}]{chi2024control}
Z.~Chi, S.~Lee, H.~Yang, E.~Dolan, C.~Safeer, J.~Ingla-Ayn{\'e}s, F.~Herling,
  N.~Ontoso, B.~Mart{\'\i}n-Garc{\'\i}a, M.~Gobbi \emph{et~al.}, \emph{Adv.
  Mater.}, 2024, \textbf{36}, 2310768\relax
\mciteBstWouldAddEndPuncttrue
\mciteSetBstMidEndSepPunct{\mcitedefaultmidpunct}
{\mcitedefaultendpunct}{\mcitedefaultseppunct}\relax
\EndOfBibitem
\bibitem[Yang \emph{et~al.}(2024)Yang, Chi, Avedissian, Dolan, Karuppasamy,
  Mart{\'\i}n-Garc{\'\i}a, Gobbi, Sofer, Hueso, and Casanova]{yang2024gate}
H.~Yang, Z.~Chi, G.~Avedissian, E.~Dolan, M.~Karuppasamy,
  B.~Mart{\'\i}n-Garc{\'\i}a, M.~Gobbi, Z.~Sofer, L.~E. Hueso and F.~Casanova,
  \emph{Adv. Func. Mater.}, 2024, \textbf{34}, 2404872\relax
\mciteBstWouldAddEndPuncttrue
\mciteSetBstMidEndSepPunct{\mcitedefaultmidpunct}
{\mcitedefaultendpunct}{\mcitedefaultseppunct}\relax
\EndOfBibitem
\bibitem[Yang \emph{et~al.}(2024)Yang, Mart{\'\i}n-Garc{\'\i}a, Kim{\'a}k,
  Schmoranzerov{\'a}, Dolan, Chi, Gobbi, N{\v{e}}mec, Hueso, and
  Casanova]{yang2024twist}
H.~Yang, B.~Mart{\'\i}n-Garc{\'\i}a, J.~Kim{\'a}k, E.~Schmoranzerov{\'a},
  E.~Dolan, Z.~Chi, M.~Gobbi, P.~N{\v{e}}mec, L.~E. Hueso and F.~Casanova,
  \emph{Nat. Mater.}, 2024, \textbf{23}, 1502--1508\relax
\mciteBstWouldAddEndPuncttrue
\mciteSetBstMidEndSepPunct{\mcitedefaultmidpunct}
{\mcitedefaultendpunct}{\mcitedefaultseppunct}\relax
\EndOfBibitem
\bibitem[Li \emph{et~al.}(2022)Li, Yao, Wu, Hu, Gao, Wan, and
  Liu]{li2022designing}
J.~Li, Q.~Yao, L.~Wu, Z.~Hu, B.~Gao, X.~Wan and Q.~Liu, \emph{Nat. Commun.},
  2022, \textbf{13}, 919\relax
\mciteBstWouldAddEndPuncttrue
\mciteSetBstMidEndSepPunct{\mcitedefaultmidpunct}
{\mcitedefaultendpunct}{\mcitedefaultseppunct}\relax
\EndOfBibitem
\bibitem[Chen \emph{et~al.}(2022)Chen, Liu, Lu, Zhao, Hu, Ren, and
  Yuan]{chen2022intrinsic}
H.~Chen, R.~Liu, J.~Lu, X.~Zhao, G.~Hu, J.~Ren and X.~Yuan, \emph{J. Phys.
  Chem. Lett.}, 2022, \textbf{13}, 10297--10304\relax
\mciteBstWouldAddEndPuncttrue
\mciteSetBstMidEndSepPunct{\mcitedefaultmidpunct}
{\mcitedefaultendpunct}{\mcitedefaultseppunct}\relax
\EndOfBibitem
\bibitem[Yu \emph{et~al.}(2021)Yu, Zhou, Zhang, and Chang]{yu2021spin}
S.-B. Yu, M.~Zhou, D.~Zhang and K.~Chang, \emph{Phys. Rev. B}, 2021,
  \textbf{104}, 075435\relax
\mciteBstWouldAddEndPuncttrue
\mciteSetBstMidEndSepPunct{\mcitedefaultmidpunct}
{\mcitedefaultendpunct}{\mcitedefaultseppunct}\relax
\EndOfBibitem
\bibitem[Manchon \emph{et~al.}(2015)Manchon, Koo, Nitta, Frolov, and
  Duine]{manchon2015new}
A.~Manchon, H.~C. Koo, J.~Nitta, S.~M. Frolov and R.~A. Duine, \emph{Nat.
  Mater.}, 2015, \textbf{14}, 871--882\relax
\mciteBstWouldAddEndPuncttrue
\mciteSetBstMidEndSepPunct{\mcitedefaultmidpunct}
{\mcitedefaultendpunct}{\mcitedefaultseppunct}\relax
\EndOfBibitem
\bibitem[Kapri \emph{et~al.}(2021)Kapri, Dey, and Ghosh]{kapri2021role}
P.~Kapri, B.~Dey and T.~K. Ghosh, \emph{Phys. Rev. B}, 2021, \textbf{103},
  165401\relax
\mciteBstWouldAddEndPuncttrue
\mciteSetBstMidEndSepPunct{\mcitedefaultmidpunct}
{\mcitedefaultendpunct}{\mcitedefaultseppunct}\relax
\EndOfBibitem
\bibitem[Chen \emph{et~al.}(2022)Chen, Zeng, Lv, Guo, Chen, and
  Geng]{chen2022large}
S.~Chen, Z.~Zeng, B.~Lv, S.~Guo, X.~Chen and H.~Geng, \emph{Phys. Rev. B},
  2022, \textbf{106}, 115307\relax
\mciteBstWouldAddEndPuncttrue
\mciteSetBstMidEndSepPunct{\mcitedefaultmidpunct}
{\mcitedefaultendpunct}{\mcitedefaultseppunct}\relax
\EndOfBibitem
\bibitem[Patel \emph{et~al.}(2022)Patel, Dey, Adhikari, and
  Taraphder]{patel2022electric}
S.~Patel, U.~Dey, N.~P. Adhikari and A.~Taraphder, \emph{Phys. Rev. B}, 2022,
  \textbf{106}, 035125\relax
\mciteBstWouldAddEndPuncttrue
\mciteSetBstMidEndSepPunct{\mcitedefaultmidpunct}
{\mcitedefaultendpunct}{\mcitedefaultseppunct}\relax
\EndOfBibitem
\bibitem[Voj{\'a}cek \emph{et~al.}(2024)Voj{\'a}cek, Medina~Due{\~n}as, Li,
  Ibrahim, Manchon, Roche, Chshiev, and Garc{\'\i}a]{vojacek2024field}
L.~Voj{\'a}cek, J.~Medina~Due{\~n}as, J.~Li, F.~Ibrahim, A.~Manchon, S.~Roche,
  M.~Chshiev and J.~H. Garc{\'\i}a, \emph{Nano Lett.}, 2024, \textbf{24},
  11889--11894\relax
\mciteBstWouldAddEndPuncttrue
\mciteSetBstMidEndSepPunct{\mcitedefaultmidpunct}
{\mcitedefaultendpunct}{\mcitedefaultseppunct}\relax
\EndOfBibitem
\bibitem[Yao \emph{et~al.}(2017)Yao, Cai, Tong, Gong, Wang, Wan, Duan, and
  Chu]{yao2017manipulation}
Q.-F. Yao, J.~Cai, W.-Y. Tong, S.-J. Gong, J.-Q. Wang, X.~Wan, C.-G. Duan and
  J.~Chu, \emph{Phys. Rev. B}, 2017, \textbf{95}, 165401\relax
\mciteBstWouldAddEndPuncttrue
\mciteSetBstMidEndSepPunct{\mcitedefaultmidpunct}
{\mcitedefaultendpunct}{\mcitedefaultseppunct}\relax
\EndOfBibitem
\bibitem[Chakraborty and Raj(2023)]{chakraborty2023anisotropic}
S.~Chakraborty and S.~Raj, \emph{Phys. Rev. B}, 2023, \textbf{107},
  035420\relax
\mciteBstWouldAddEndPuncttrue
\mciteSetBstMidEndSepPunct{\mcitedefaultmidpunct}
{\mcitedefaultendpunct}{\mcitedefaultseppunct}\relax
\EndOfBibitem
\bibitem[Yang \emph{et~al.}(2023)Yang, Li, Liu, and Bai]{yang2023tunable}
C.~Yang, J.~Li, X.~Liu and C.~Bai, \emph{Phys. Chem. Chem. Phys.}, 2023,
  \textbf{25}, 28796--28806\relax
\mciteBstWouldAddEndPuncttrue
\mciteSetBstMidEndSepPunct{\mcitedefaultmidpunct}
{\mcitedefaultendpunct}{\mcitedefaultseppunct}\relax
\EndOfBibitem
\bibitem[Thiruvengadam \emph{et~al.}(2022)Thiruvengadam, Mishra, Mohanty, and
  Bedanta]{thiruvengadam2022anisotropy}
V.~Thiruvengadam, A.~Mishra, S.~Mohanty and S.~Bedanta, \emph{ACS Appl. Nano
  Mater.}, 2022, \textbf{5}, 10645--10651\relax
\mciteBstWouldAddEndPuncttrue
\mciteSetBstMidEndSepPunct{\mcitedefaultmidpunct}
{\mcitedefaultendpunct}{\mcitedefaultseppunct}\relax
\EndOfBibitem
\bibitem[Mishra \emph{et~al.}(2024)Mishra, Das, Chhatoi, Dash, Sahoo, Rathore,
  Cha, Lee, Bhattacharjee, and Bedanta]{mishra2024preserving}
A.~Mishra, P.~Das, R.~Chhatoi, S.~Dash, S.~Sahoo, K.~S. Rathore, P.-R. Cha,
  S.-C. Lee, S.~Bhattacharjee and S.~Bedanta, \emph{arXiv preprint
  arXiv:2411.18582}, 2024\relax
\mciteBstWouldAddEndPuncttrue
\mciteSetBstMidEndSepPunct{\mcitedefaultmidpunct}
{\mcitedefaultendpunct}{\mcitedefaultseppunct}\relax
\EndOfBibitem
\bibitem[Wang \emph{et~al.}(2024)Wang, Meng, Lin, Xu, Ma, Kan, Chen, Huang,
  Chen, Yue,\emph{et~al.}]{wang2024origin}
H.~Wang, J.~Meng, J.~Lin, B.~Xu, H.~Ma, Y.~Kan, R.~Chen, L.~Huang, Y.~Chen,
  F.~Yue \emph{et~al.}, \emph{Nat. Commun.}, 2024, \textbf{15}, 4362\relax
\mciteBstWouldAddEndPuncttrue
\mciteSetBstMidEndSepPunct{\mcitedefaultmidpunct}
{\mcitedefaultendpunct}{\mcitedefaultseppunct}\relax
\EndOfBibitem
\bibitem[Knispel \emph{et~al.}(2024)Knispel, Berges, Schobert, van Loon, Jolie,
  Wehling, Michely, and Fischer]{knispel2024unconventional}
T.~Knispel, J.~Berges, A.~Schobert, E.~G. van Loon, W.~Jolie, T.~Wehling,
  T.~Michely and J.~Fischer, \emph{Nano Lett.}, 2024, \textbf{24},
  1045--1051\relax
\mciteBstWouldAddEndPuncttrue
\mciteSetBstMidEndSepPunct{\mcitedefaultmidpunct}
{\mcitedefaultendpunct}{\mcitedefaultseppunct}\relax
\EndOfBibitem
\bibitem[Xi \emph{et~al.}(2015)Xi, Zhao, Wang, Berger, Forr{\'o}, Shan, and
  Mak]{xi2015strongly}
X.~Xi, L.~Zhao, Z.~Wang, H.~Berger, L.~Forr{\'o}, J.~Shan and K.~F. Mak,
  \emph{Nat. Nanotechnol.}, 2015, \textbf{10}, 765--769\relax
\mciteBstWouldAddEndPuncttrue
\mciteSetBstMidEndSepPunct{\mcitedefaultmidpunct}
{\mcitedefaultendpunct}{\mcitedefaultseppunct}\relax
\EndOfBibitem
\bibitem[Nakata \emph{et~al.}(2021)Nakata, Sugawara, Chainani, Oka, Bao, Zhou,
  Chuang, Cheng, Kawakami, Saruta,\emph{et~al.}]{nakata2021robust}
Y.~Nakata, K.~Sugawara, A.~Chainani, H.~Oka, C.~Bao, S.~Zhou, P.-Y. Chuang,
  C.-M. Cheng, T.~Kawakami, Y.~Saruta \emph{et~al.}, \emph{Nat. Commun.}, 2021,
  \textbf{12}, 5873\relax
\mciteBstWouldAddEndPuncttrue
\mciteSetBstMidEndSepPunct{\mcitedefaultmidpunct}
{\mcitedefaultendpunct}{\mcitedefaultseppunct}\relax
\EndOfBibitem
\bibitem[Hamill \emph{et~al.}(2021)Hamill, Heischmidt, Sohn, Shaffer, Tsai,
  Zhang, Xi, Suslov, Berger, Forr{\'o},\emph{et~al.}]{hamill2021two}
A.~Hamill, B.~Heischmidt, E.~Sohn, D.~Shaffer, K.-T. Tsai, X.~Zhang, X.~Xi,
  A.~Suslov, H.~Berger, L.~Forr{\'o} \emph{et~al.}, \emph{Nat. Phys.}, 2021,
  \textbf{17}, 949--954\relax
\mciteBstWouldAddEndPuncttrue
\mciteSetBstMidEndSepPunct{\mcitedefaultmidpunct}
{\mcitedefaultendpunct}{\mcitedefaultseppunct}\relax
\EndOfBibitem
\bibitem[Wickramaratne \emph{et~al.}(2020)Wickramaratne, Khmelevskyi,
  Agterberg, and Mazin]{wickramaratne2020ising}
D.~Wickramaratne, S.~Khmelevskyi, D.~F. Agterberg and I.~Mazin, \emph{Phys.
  Rev. X}, 2020, \textbf{10}, 041003\relax
\mciteBstWouldAddEndPuncttrue
\mciteSetBstMidEndSepPunct{\mcitedefaultmidpunct}
{\mcitedefaultendpunct}{\mcitedefaultseppunct}\relax
\EndOfBibitem
\bibitem[Sohn \emph{et~al.}(2018)Sohn, Xi, He, Jiang, Wang, Kang, Park, Berger,
  Forr{\'o}, Law,\emph{et~al.}]{sohn2018unusual}
E.~Sohn, X.~Xi, W.-Y. He, S.~Jiang, Z.~Wang, K.~Kang, J.-H. Park, H.~Berger,
  L.~Forr{\'o}, K.~T. Law \emph{et~al.}, \emph{Nat. Mater.}, 2018, \textbf{17},
  504--508\relax
\mciteBstWouldAddEndPuncttrue
\mciteSetBstMidEndSepPunct{\mcitedefaultmidpunct}
{\mcitedefaultendpunct}{\mcitedefaultseppunct}\relax
\EndOfBibitem
\bibitem[Xi \emph{et~al.}(2016)Xi, Wang, Zhao, Park, Law, Berger, Forr{\'o},
  Shan, and Mak]{xi2016ising}
X.~Xi, Z.~Wang, W.~Zhao, J.-H. Park, K.~T. Law, H.~Berger, L.~Forr{\'o},
  J.~Shan and K.~F. Mak, \emph{Nat. Phys.}, 2016, \textbf{12}, 139--143\relax
\mciteBstWouldAddEndPuncttrue
\mciteSetBstMidEndSepPunct{\mcitedefaultmidpunct}
{\mcitedefaultendpunct}{\mcitedefaultseppunct}\relax
\EndOfBibitem
\bibitem[Xu \emph{et~al.}(2018)Xu, Wang, Chen, Fl{\"o}totto, Hlevyack, Lin,
  Bian, Mo, and Chiang]{xu2018experimental}
C.-Z. Xu, X.~Wang, P.~Chen, D.~Fl{\"o}totto, J.~A. Hlevyack, M.-K. Lin,
  G.~Bian, S.-K. Mo and T.-C. Chiang, \emph{Phys. Rev. Mater.}, 2018,
  \textbf{2}, 064002\relax
\mciteBstWouldAddEndPuncttrue
\mciteSetBstMidEndSepPunct{\mcitedefaultmidpunct}
{\mcitedefaultendpunct}{\mcitedefaultseppunct}\relax
\EndOfBibitem
\bibitem[De~la Barrera \emph{et~al.}(2018)De~la Barrera, Sinko, Gopalan,
  Sivadas, Seyler, Watanabe, Taniguchi, Tsen, Xu,
  Xiao,\emph{et~al.}]{de2018tuning}
S.~C. De~la Barrera, M.~R. Sinko, D.~P. Gopalan, N.~Sivadas, K.~L. Seyler,
  K.~Watanabe, T.~Taniguchi, A.~W. Tsen, X.~Xu, D.~Xiao \emph{et~al.},
  \emph{Nat. Commun.}, 2018, \textbf{9}, 1427\relax
\mciteBstWouldAddEndPuncttrue
\mciteSetBstMidEndSepPunct{\mcitedefaultmidpunct}
{\mcitedefaultendpunct}{\mcitedefaultseppunct}\relax
\EndOfBibitem
\bibitem[Habara and Wakabayashi(2021)]{habara2021optically}
R.~Habara and K.~Wakabayashi, \emph{Phys. Rev. B}, 2021, \textbf{103},
  L161410\relax
\mciteBstWouldAddEndPuncttrue
\mciteSetBstMidEndSepPunct{\mcitedefaultmidpunct}
{\mcitedefaultendpunct}{\mcitedefaultseppunct}\relax
\EndOfBibitem
\bibitem[Habara and Wakabayashi(2022)]{habara2022nonlinear}
R.~Habara and K.~Wakabayashi, \emph{Phys. Rev. Res.}, 2022, \textbf{4},
  013219\relax
\mciteBstWouldAddEndPuncttrue
\mciteSetBstMidEndSepPunct{\mcitedefaultmidpunct}
{\mcitedefaultendpunct}{\mcitedefaultseppunct}\relax
\EndOfBibitem
\bibitem[Adhikary and Wakabayashi(2024)]{adhikary2024optically}
S.~Adhikary and K.~Wakabayashi, \emph{J. Phys. Chem. C}, 2024, \textbf{128},
  14514--14521\relax
\mciteBstWouldAddEndPuncttrue
\mciteSetBstMidEndSepPunct{\mcitedefaultmidpunct}
{\mcitedefaultendpunct}{\mcitedefaultseppunct}\relax
\EndOfBibitem
\bibitem[Wu \emph{et~al.}(2025)Wu, Sayyad, Sailus, Dey, Xie, Hays, Kopaczek,
  Ou, Susarla, Esqueda,\emph{et~al.}]{wu2025metallic}
C.-L. Wu, M.~Y. Sayyad, R.~E. Sailus, D.~Dey, J.~Xie, P.~Hays, J.~Kopaczek,
  Y.~Ou, S.~Susarla, I.~S. Esqueda \emph{et~al.}, \emph{Nanoscale}, 2025,
  \textbf{17}, 7801--7812\relax
\mciteBstWouldAddEndPuncttrue
\mciteSetBstMidEndSepPunct{\mcitedefaultmidpunct}
{\mcitedefaultendpunct}{\mcitedefaultseppunct}\relax
\EndOfBibitem
\bibitem[G{\"u}ller \emph{et~al.}(2016)G{\"u}ller, Vildosola, and
  Llois]{guller2016spin}
F.~G{\"u}ller, V.~L. Vildosola and A.~M. Llois, \emph{Phys. Rev. B}, 2016,
  \textbf{93}, 094434\relax
\mciteBstWouldAddEndPuncttrue
\mciteSetBstMidEndSepPunct{\mcitedefaultmidpunct}
{\mcitedefaultendpunct}{\mcitedefaultseppunct}\relax
\EndOfBibitem
\bibitem[Giannozzi \emph{et~al.}(2017)Giannozzi, Andreussi, Brumme, Bunau,
  Nardelli, Calandra, Car, Cavazzoni, Ceresoli,
  Cococcioni,\emph{et~al.}]{giannozzi2017advanced}
P.~Giannozzi, O.~Andreussi, T.~Brumme, O.~Bunau, M.~B. Nardelli, M.~Calandra,
  R.~Car, C.~Cavazzoni, D.~Ceresoli, M.~Cococcioni \emph{et~al.}, \emph{J.
  Phys. Condens. Matter}, 2017, \textbf{29}, 465901\relax
\mciteBstWouldAddEndPuncttrue
\mciteSetBstMidEndSepPunct{\mcitedefaultmidpunct}
{\mcitedefaultendpunct}{\mcitedefaultseppunct}\relax
\EndOfBibitem
\bibitem[Perdew \emph{et~al.}(1996)Perdew, Burke, and
  Ernzerhof]{perdew1996generalized}
J.~P. Perdew, K.~Burke and M.~Ernzerhof, \emph{Phys. Rev. Lett.}, 1996,
  \textbf{77}, 3865\relax
\mciteBstWouldAddEndPuncttrue
\mciteSetBstMidEndSepPunct{\mcitedefaultmidpunct}
{\mcitedefaultendpunct}{\mcitedefaultseppunct}\relax
\EndOfBibitem
\bibitem[Van~Setten \emph{et~al.}(2018)Van~Setten, Giantomassi, Bousquet,
  Verstraete, Hamann, Gonze, and Rignanese]{van2018pseudodojo}
M.~J. Van~Setten, M.~Giantomassi, E.~Bousquet, M.~J. Verstraete, D.~R. Hamann,
  X.~Gonze and G.-M. Rignanese, \emph{Comput. Phys. Commun.}, 2018,
  \textbf{226}, 39--54\relax
\mciteBstWouldAddEndPuncttrue
\mciteSetBstMidEndSepPunct{\mcitedefaultmidpunct}
{\mcitedefaultendpunct}{\mcitedefaultseppunct}\relax
\EndOfBibitem
\bibitem[Mostofi \emph{et~al.}(2008)Mostofi, Yates, Lee, Souza, Vanderbilt, and
  Marzari]{mostofi2008wannier90}
A.~A. Mostofi, J.~R. Yates, Y.-S. Lee, I.~Souza, D.~Vanderbilt and N.~Marzari,
  \emph{Comput. Phys. Commun.}, 2008, \textbf{178}, 685--699\relax
\mciteBstWouldAddEndPuncttrue
\mciteSetBstMidEndSepPunct{\mcitedefaultmidpunct}
{\mcitedefaultendpunct}{\mcitedefaultseppunct}\relax
\EndOfBibitem
\bibitem[Qiao \emph{et~al.}(2018)Qiao, Zhou, Yuan, and
  Zhao]{qiao2018calculation}
J.~Qiao, J.~Zhou, Z.~Yuan and W.~Zhao, \emph{Phys. Rev. B}, 2018, \textbf{98},
  214402\relax
\mciteBstWouldAddEndPuncttrue
\mciteSetBstMidEndSepPunct{\mcitedefaultmidpunct}
{\mcitedefaultendpunct}{\mcitedefaultseppunct}\relax
\EndOfBibitem
\bibitem[Guo \emph{et~al.}(2005)Guo, Yao, and Niu]{guo2005ab}
G.~Guo, Y.~Yao and Q.~Niu, \emph{Phys. Rev. Lett.}, 2005, \textbf{94},
  226601\relax
\mciteBstWouldAddEndPuncttrue
\mciteSetBstMidEndSepPunct{\mcitedefaultmidpunct}
{\mcitedefaultendpunct}{\mcitedefaultseppunct}\relax
\EndOfBibitem
\bibitem[Yates \emph{et~al.}(2007)Yates, Wang, Vanderbilt, and
  Souza]{yates2007spectral}
J.~R. Yates, X.~Wang, D.~Vanderbilt and I.~Souza, \emph{Phys. Rev. B}, 2007,
  \textbf{75}, 195121\relax
\mciteBstWouldAddEndPuncttrue
\mciteSetBstMidEndSepPunct{\mcitedefaultmidpunct}
{\mcitedefaultendpunct}{\mcitedefaultseppunct}\relax
\EndOfBibitem
\bibitem[Cossu \emph{et~al.}(2020)Cossu, Palot{\'a}s, Sarkar, Di~Marco, and
  Akbari]{cossu2020strain}
F.~Cossu, K.~Palot{\'a}s, S.~Sarkar, I.~Di~Marco and A.~Akbari, \emph{NPG Asia
  Mater.}, 2020, \textbf{12}, 24\relax
\mciteBstWouldAddEndPuncttrue
\mciteSetBstMidEndSepPunct{\mcitedefaultmidpunct}
{\mcitedefaultendpunct}{\mcitedefaultseppunct}\relax
\EndOfBibitem
\bibitem[Ichinokura \emph{et~al.}(2019)Ichinokura, Nakata, Sugawara, Endo,
  Takayama, Takahashi, and Hasegawa]{ichinokura2019vortex}
S.~Ichinokura, Y.~Nakata, K.~Sugawara, Y.~Endo, A.~Takayama, T.~Takahashi and
  S.~Hasegawa, \emph{Phys. Rev. B}, 2019, \textbf{99}, 220501\relax
\mciteBstWouldAddEndPuncttrue
\mciteSetBstMidEndSepPunct{\mcitedefaultmidpunct}
{\mcitedefaultendpunct}{\mcitedefaultseppunct}\relax
\EndOfBibitem
\bibitem[Zhou \emph{et~al.}(2019)Zhou, Chen, Yang, Liu, and
  Ouyang]{zhou2019geometry}
W.~Zhou, J.~Chen, Z.~Yang, J.~Liu and F.~Ouyang, \emph{Phys. Rev. B}, 2019,
  \textbf{99}, 075160\relax
\mciteBstWouldAddEndPuncttrue
\mciteSetBstMidEndSepPunct{\mcitedefaultmidpunct}
{\mcitedefaultendpunct}{\mcitedefaultseppunct}\relax
\EndOfBibitem
\bibitem[Sino \emph{et~al.}(2021)Sino, Feng, Villaos, Cruzado, Huang, Hsu, and
  Chuang]{sino2021anisotropic}
P.~A.~L. Sino, L.-Y. Feng, R.~A.~B. Villaos, H.~N. Cruzado, Z.-Q. Huang, C.-H.
  Hsu and F.-C. Chuang, \emph{Nanoscale Adv.}, 2021, \textbf{3},
  6608--6616\relax
\mciteBstWouldAddEndPuncttrue
\mciteSetBstMidEndSepPunct{\mcitedefaultmidpunct}
{\mcitedefaultendpunct}{\mcitedefaultseppunct}\relax
\EndOfBibitem
\bibitem[Seemann \emph{et~al.}(2015)Seemann, K{\"o}dderitzsch, Wimmer, and
  Ebert]{seemann2015symmetry}
M.~Seemann, D.~K{\"o}dderitzsch, S.~Wimmer and H.~Ebert, \emph{Phys. Rev. B},
  2015, \textbf{92}, 155138\relax
\mciteBstWouldAddEndPuncttrue
\mciteSetBstMidEndSepPunct{\mcitedefaultmidpunct}
{\mcitedefaultendpunct}{\mcitedefaultseppunct}\relax
\EndOfBibitem
\bibitem[Xiang \emph{et~al.}(2024)Xiang, Xu, Wang, and
  Wang]{xiang2024classification}
L.~Xiang, F.~Xu, L.~Wang and J.~Wang, \emph{Front. Phys.}, 2024, \textbf{19},
  33205\relax
\mciteBstWouldAddEndPuncttrue
\mciteSetBstMidEndSepPunct{\mcitedefaultmidpunct}
{\mcitedefaultendpunct}{\mcitedefaultseppunct}\relax
\EndOfBibitem
\bibitem[Kubetzka \emph{et~al.}(2002)Kubetzka, Bode, Pietzsch, and
  Wiesendanger]{kubetzka2002spin}
A.~Kubetzka, M.~Bode, O.~Pietzsch and R.~Wiesendanger, \emph{Phys. Rev. Lett.},
  2002, \textbf{88}, 057201\relax
\mciteBstWouldAddEndPuncttrue
\mciteSetBstMidEndSepPunct{\mcitedefaultmidpunct}
{\mcitedefaultendpunct}{\mcitedefaultseppunct}\relax
\EndOfBibitem
\bibitem[Kawaguchi \emph{et~al.}(2023)Kawaguchi, Kuroda, Zhao, Tani, Harasawa,
  Fukushima, Tanaka, Noguchi, Iimori, Yaji,\emph{et~al.}]{kawaguchi2023time}
K.~Kawaguchi, K.~Kuroda, Z.~Zhao, S.~Tani, A.~Harasawa, Y.~Fukushima,
  H.~Tanaka, R.~Noguchi, T.~Iimori, K.~Yaji \emph{et~al.}, \emph{Rev. Sci.
  Instrum.}, 2023, \textbf{94}, 083902\relax
\mciteBstWouldAddEndPuncttrue
\mciteSetBstMidEndSepPunct{\mcitedefaultmidpunct}
{\mcitedefaultendpunct}{\mcitedefaultseppunct}\relax
\EndOfBibitem
\bibitem[Farooq \emph{et~al.}(2022)Farooq, Xian, and Huang]{farooq2022spin}
M.~U. Farooq, L.~Xian and L.~Huang, \emph{Phys. Rev. B}, 2022, \textbf{105},
  245405\relax
\mciteBstWouldAddEndPuncttrue
\mciteSetBstMidEndSepPunct{\mcitedefaultmidpunct}
{\mcitedefaultendpunct}{\mcitedefaultseppunct}\relax
\EndOfBibitem
\bibitem[Gong \emph{et~al.}(2024)Gong, Li, Wang, and Zhang]{gong2024large}
L.~Gong, Y.~Li, H.~Wang and H.~Zhang, \emph{Phys. Rev. B}, 2024, \textbf{109},
  045124\relax
\mciteBstWouldAddEndPuncttrue
\mciteSetBstMidEndSepPunct{\mcitedefaultmidpunct}
{\mcitedefaultendpunct}{\mcitedefaultseppunct}\relax
\EndOfBibitem
\bibitem[Homes \emph{et~al.}(2013)Homes, Tu, Li, Gu, and
  Akrap]{homes2013optical}
C.~Homes, J.~Tu, J.~Li, G.~Gu and A.~Akrap, \emph{Sci. Rep.}, 2013, \textbf{3},
  3446\relax
\mciteBstWouldAddEndPuncttrue
\mciteSetBstMidEndSepPunct{\mcitedefaultmidpunct}
{\mcitedefaultendpunct}{\mcitedefaultseppunct}\relax
\EndOfBibitem
\bibitem[Schindlmayr(2009)]{schindlmayr2009optical}
A.~Schindlmayr, AIP Conf. Proc., 2009, pp. 157--159\relax
\mciteBstWouldAddEndPuncttrue
\mciteSetBstMidEndSepPunct{\mcitedefaultmidpunct}
{\mcitedefaultendpunct}{\mcitedefaultseppunct}\relax
\EndOfBibitem
\end{mcitethebibliography}
\bibliographystyle{rsc} 

\end{document}